\documentclass[aps, prx, superscriptaddress, reprint]{revtex4-1}
\usepackage{graphicx, amsmath, amsthm, amssymb, epstopdf}
\DeclareMathOperator*{\argmin}{arg\,min}
\usepackage[ruled, noline]{algorithm2e}
\begin{document}
\title{Using photoelectron spectroscopy to measure resonant inelastic X-ray scattering: A computational investigation}
\author{Daniel J. Higley}
\email{dhigley@slac.stanford.edu}
\affiliation{SLAC National Accelerator Laboratory, 2575 Sand Hill Road, Menlo
Park, California 94025, USA}
\author{Hirohito Ogasawara}
\affiliation{SLAC National Accelerator Laboratory, 2575 Sand Hill Road, Menlo
  Park, California 94025, USA}
\author{Sioan Zohar}
\affiliation{SLAC National Accelerator Laboratory, 2575 Sand Hill Road, Menlo
  Park, California 94025, USA}
\author{Georgi L. Dakovski}
\affiliation{SLAC National Accelerator Laboratory, 2575 Sand Hill Road, Menlo
Park, California 94025, USA}
\begin{abstract}
  Resonant inelastic X-ray scattering (RIXS) has become an important scientific tool. Nonetheless, conventional high-resolution RIXS measurements ($<100$ meV), especially in the soft x-ray range, require large and low-throughput grating  spectrometers that limits measurement accuracy and simplicity. Here, we computationally investigate the performance of a different method for measuring RIXS, Photoelectron Spectrometry for Analysis of X-rays (PAX). This method transforms the X-ray measurement problem of RIXS to an electron measurement problem, enabling use of compact, high-throughput electron spectrometers. In PAX, X-rays to be measured are incident on a converter material and the energy distribution of the resultant photoelectrons, the PAX spectrum, is measured with an electron spectrometer. The incident X-ray spectrum is then estimated through a deconvolution algorithm that leverages concepts from machine learning. We investigate a few example PAX cases. Using the 3d levels of Ag as a converter material, and with $10^5$ detected electrons, we accurately estimate features with 100s of meV width in a model RIXS spectrum. Using a sharp Fermi edge to encode RIXS spectra, we accurately distinguish 100 meV FWHM peaks separated by 45 meV with $10^7$ electrons detected that were photoemitted from within 0.4 eV of the Fermi level.
\end{abstract}

\maketitle

\section{Introduction}

Resonant Inelastic X-Ray Scattering (RIXS) has emerged as a powerful technique to study elementary excitations \cite{ament2011resonant}. RIXS probes excitations via core-valence transitions with element-specific energies, which allows one to tune the elemental locations being probed through the incident X-ray photon energy. Because of the large momentum of X-rays, RIXS is able to probe the dispersion of elementary excitations in solids, unlike lower energy optical photons. Further, in contrast to other X-ray-based spectroscopies, the energy resolution of RIXS is not limited by short core hole lifetimes. Because of these strengths, much effort has been devoted to developing RIXS capabilities. In the soft x-ray range, X-ray grating and synchrotron light source development has enabled RIXS measurements with energy resolution better than 100 meV \cite{brookes2018beamline}. Having these high energy resolutions is particularly important for studies of solids where characteristic energies of many important excitations are around 100 meV or less \cite{ament2011resonant, chaix2017dispersive}. Leveraging these capabilities, RIXS studies have given new insights in wide-ranging topics including solid state physics \cite{ament2011resonant, le2011intense, schlappa2012spin, chaix2017dispersive}, nanoparticles \cite{liu2017situ}, interfaces \cite{rajasekaran2012probing}, batteries \cite{house2020superstructure, firouzi2018monovalent}, liquids \cite{wernet2015orbital} and gases \cite{hennies2010resonant}.

 Despite these strengths and much instrumentation development \cite{brookes2018beamline}, grating spectrometers only detect one X-ray photon for every $\approx10^6$ X-ray photons incident on a sample \cite{ghiringhelli2013magnetic, dakovski2017novel}. Thus, RIXS is a very photon hungry technique. In addition, the X-ray flux incident on a sample is constrained by radiation damage and X-ray source output. This limits achievable count rate, and thus accuracy, of RIXS measurements. Work has been done towards improving this. Transition edge sensors \cite{uhlig2015high} and off-axis zone plates \cite{marschall2017transmission} can make more accurate RIXS measurements than traditional instrumentation in certain cases, but so-far demonstrated resolutions are significantly more than 0.5 eV. More information could be gleaned from RIXS if one could make faster and more accurate measurements that maintained hundreds of meV or better resolution. For example, high resolution time-resolved RIXS measurements probe dynamics of elementary excitations and can discriminate well between different states that occur in sample evolution \cite{wernet2015orbital}, but are very challenging due to their requirement of recording many accurate spectra. 

Coinciding with the rise and development of RIXS, the capabilities of electron spectrometers have also greatly increased \cite{damascelli2003angle}. Photoelectron spectrometers can now have energy resolutions better than 20 meV for 500 eV kinetic energy electrons \cite{seidel2017advances}. The collection efficiency, and thus achievable signal-to-noise ratio with a given number of particles emitted from a sample, can be much higher for electron spectrometers than X-ray spectrometers with comparable resolutions. Further, the footprint of electron spectrometers with 10s of meV resolution when measuring several hundred eV electrons (few m$^2$) is far smaller than current X-ray spectrometers with comparable resolutions for several hundred eV photons (few hundred m$^2$) \cite{brookes2018beamline}. These features are largely due to the ease with which electrons can be manipulated, owing to their charge, in comparison with neutral X-rays. Thus if one can transform the X-ray measurement problem of RIXS to an electron measurement problem, then there could be large gains in count rates, as well as instrumentation compactness and ease of implementation.

\begin{figure*}
    \centering
    \includegraphics{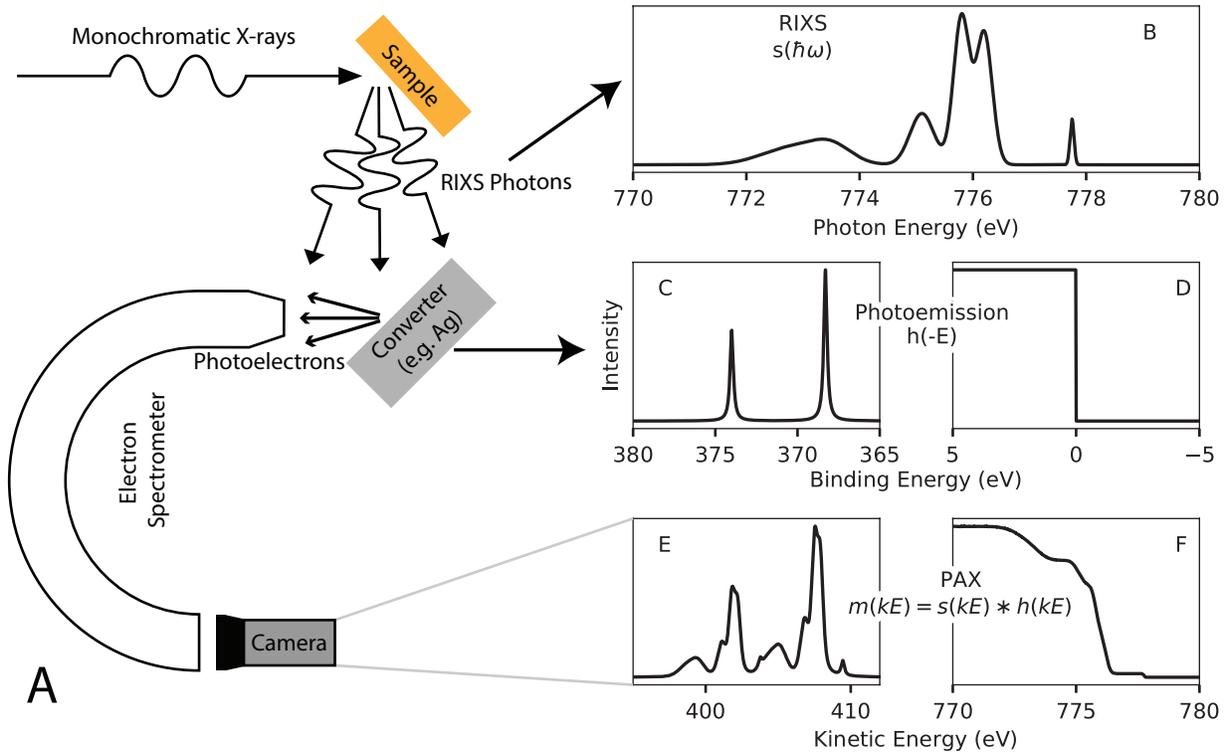}
    \caption{Concept of PAX. (A) Experimental schematic. The measured PAX spectrum (E, F) is approximately the convolution of the X-ray spectrum incident on the converter material (B) and the photoemission spectrum of the converter material (C, D). The X-ray spectrum (B) is a model of a RIXS spectrum recorded at the Co L$_3$ edge (778 eV incident X-ray photon energy). The photoemission spectra model photoemission from the Ag 3d levels (C) and a sharp Fermi edge (D). The PAX spectrum shown in in (E) is calculated with the photoemission spectrum of (B), while (F) is calculated using (D).}
    \label{fig:overview}
\end{figure*}

Recently, Dakovski \textit{et al.} \cite{dakovski2017novel} proposed doing exactly that to measure Resonant Inelastic X-ray Scattering (RIXS) with hundreds of meV or better resolution through Photoelectron Spectrometry for Analysis of X-rays (PAX) \cite{krause1965determination, ebel1975use}. Fig. \ref{fig:overview} gives an overview of this technique, which makes use of sharp photoemission features that occur in the photoemission spectra of materials such as Ag, Au, Pt or Al when measured with monochromatic incident X-ray radiation. Fig. \ref{fig:overview}A shows an experimental schematic. X-rays to be measured (Fig. \ref{fig:overview}B) are incident on a converter system, where absorption of the X-rays generates photoelectrons. The converter material is assumed to give some photoemission spectrum when measured with monochromatic X-ray radiation, $xps(BE)$, as a function of binding energy, $BE$ (Fig. \ref{fig:overview}C shows an example Ag 3d photoemission spectrum, while Fig. \ref{fig:overview}D shows an example sharp Fermi edge). The emitted photoelectrons are then detected with a photoelectron spectrometer. We call the resultant electron spectrum the PAX spectrum (Fig. \ref{fig:overview}E or F). The expected shape of the PAX spectrum is given by a convolution with $h(E)=xps(-E)$ acting as an impulse response function. Convolving the spectrum of X-rays incident on the converter material (Fig. \ref{fig:overview}B) with $h(E)$ (Fig. \ref{fig:overview}C or D) approximately gives the expected value of the PAX spectrum (Fig. \ref{fig:overview}E or F),
\begin{equation}\label{eq:pax_definition_continuous}
E\{m(kE)\} = \int_{0}^{\infty}s(\hbar\omega)h(kE-\hbar\omega)d\hbar\omega,
\end{equation}
where we only integrate over physically realistic positive photon energies. Here, $kE$ is electron kinetic energy, $s(\hbar\omega)$ is the X-ray spectrum incident on the converter material as a function of photon energy, $\hbar\omega$, and $m(kE)$ is a measured PAX spectrum. 

Given a PAX spectrum, the ground truth X-ray spectrum, $s(\hbar\omega)$, can be estimated directly through deconvolution or in a parameterized form such as a sum of peaks. (The ground truth X-ray spectrum is the X-ray spectrum that would be measured without noise and with perfect resolution.) The decomposition of RIXS spectra into a sum of peaks is a natural method already widely used for traditionally recorded RIXS spectra as the parameters of these peaks can be directly linked to physical characteristics of the matter under study \cite{ament2011resonant}. For more complex RIXS spectra, or cases where less is known about the form of a RIXS spectrum before measurement, the more general case of deconvolution may be more appropriate or would be a first step in further decomposition of a RIXS spectrum into elementary features.

While Dakovski \textit{et al.} \cite{dakovski2017novel} demonstrated the possibility of recording PAX spectra for RIXS and estimating the corresponding RIXS spectra as a sum of peaks, a general algorithm for faithfully reconstructing X-ray spectra from PAX measurements, and a quantitative assessment of the potential of PAX for measurement of X-ray spectra are needed. Here, we fill these gaps by proposing an algorithm for analyzing PAX data using methods from statistical data analysis and machine learning \cite{fister2007deconvolving, bertero2009image, james2013introduction}, then characterizing the performance of PAX in the estimation of model RIXS spectra using the Ag 3d levels or a sharp Fermi edge as a model converter.

The rest of this report is organized as follows. In section \ref{sec:converter}, we discuss the considerations for choosing a converter material for PAX, and explain why the Ag 3d lines and a sharp Fermi edge are compelling cases. In section \ref{sec:deconv} we describe and discuss the deconvolution algorithm we used to estimate RIXS spectra from simulated PAX spectra. In section \ref{sec:results} we show the simulated performance of PAX in estimating model RIXS features. We find that, using the Ag 3d levels as a photoemission converter, PAX can accurately estimate the width of few hundred meV features when 10$^5$ electrons are detected in the measured PAX spectrum. Using a sharp Fermi edge photoemission converter shows promise for estimating finer features. Finally, in section \ref{sec:conclusions} we conclude and give an outlook for future investigations.

\section{Choice of Converter Material}\label{sec:converter}

\begin{figure}[htb]
    \centering
    \includegraphics{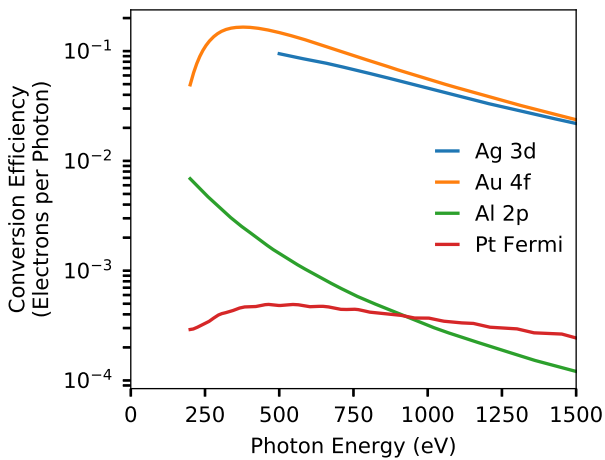}
    \caption{Photoemission quantum efficiency of some electronic subshells.}
    \label{fig:efficiency}
\end{figure}

\begin{table}[tb]
\centering
\begin{tabular}{|p{3cm}|p{3cm}|}
\hline
Subshell & Description \\
\hline
Ag 3d & Two 233 meV FWHM peaks\cite{panaccione2005high} \\
Au 4f & Two 335 meV FWHM peaks\cite{takata2005development} \\
Al 2p & Two 60 meV FWHM peaks\cite{borg2004experimental} \\
Pt Valence & Sharp Fermi edge \\
\hline
\end{tabular}
\caption{Description of photoemission from some electronic subshells of solids.}
\label{table:conversion}
\end{table}

The converter material plays a key role in PAX measurements. Converter materials with high conversion efficiency and narrow photoemssion lines are desirable. Higher conversion efficiencies give higher numbers of detected electrons and thus, potentially, higher signal-to-noise ratios. Narrow photoemission lines enable retrieval of narrow X-ray spectral features with a reasonable number of detected electrons. For thick converter materials with near normal incidence of X-rays and emission of photoelectrons, the conversion efficiency of X-rays to photoelectrons can be approximated as \cite{henke1972ultrasoft}
\begin{equation}\label{eq:conversion_efficiency}
    \epsilon = \tau\rho\lambda_e,
\end{equation}
where $\tau_q$ is the effective ionzation cross-section for the creation of photoelectrons from subshell $q$, $\rho$ is the number per unit volume of atoms within the sample which can emit photoelectrons from the subshell used for PAX, and $\lambda_e$ is the electron mean free path at the kinetic energy of the relevant photoelectrons.

In Fig. \ref{fig:efficiency} we show the conversion efficiency estimated with Eq. \ref{eq:conversion_efficiency} for some promising cases. For these calculations subshell photoemission cross sections were taken from \cite{yeh1985atomic}, electron mean free paths for Ag and Au were taken from \cite{tanuma2002experimental}, electron mean free paths for Al and Pt were taken from \cite{shinotsuka2015calculations}, and atoms per unit volume were calculated from values in \cite{rumble2019crc}. From Fig. \ref{fig:efficiency}, we see that, in the soft X-ray range, conversion efficiencies of nearly ten percent with photoemission linewidths of a few hundred meV are possible using the Au 4f or Ag 3d lines. Narrower photoemission features are available at the expense of a reduced conversion efficiency (such as Al 2p and Pt Fermi level photoemission shown in Fig. \ref{fig:efficiency}).

\section{An Algorithm for Deconvolving PAX Spectra}\label{sec:deconv}
In principle, a similar convolution equation as Eq. \ref{eq:pax_definition_continuous} describes the measured signal in many X-ray spectroscopies. The signal of interest, such as a RIXS spectrum or X-ray absorption spectrum, is convolved with an instrument response function and an intrinsic broadening function to give the measured spectrum. Thus, more accurate X-ray spectra can often be retrieved through deconvolution of measured results \cite{ebel1975deconvolution, fister2007deconvolving, laverock2011maximum}. It is not typical to analyze such spectra using deconvolution however. This is because these convolutions only broaden the measured spectra, and measured spectra are still interpretable as a simple blurring of the true spectrum. For PAX, however, the measured spectrum is typically convolved with a more complicated function than a single peak. The converter material photoelectron spectrum could consist of, for example, two narrow peaks and a non-uniform background, as is common for core levels. It may not be easy to infer the original X-ray spectrum from the measured PAX spectrum in these cases. Thus, while deconvolution is an optional step in traditional X-ray spectroscopies, it is  important for PAX measurements.

\subsection{Model of PAX Spectra}

Eq. \ref{eq:pax_definition_continuous} gives the expected value of a PAX measurement in the case that the PAX spectrum is measured at every electron kinetic energy. In reality, the measured PAX spectrum, $m[kE]$ is discrete with each measured point integrating over a range of electron kinetic energies. Thus, the expected value of the measured PAX spectrum is approximately given by a discrete convolution,
\begin{equation}\label{eq:full_discrete_pax}
    E\{m[kE]\} = \sum_{\hbar\omega = 0}^{\infty} s[\hbar\omega]h[kE-\hbar\omega],
\end{equation}
where $s[\hbar\omega]$ and $h[-BE]$ are the discretized versions of the X-ray spectrum incident on the converter material and the photoemission spectrum of the converter material. 

Eq. \ref{eq:full_discrete_pax} extends over an infinite range, but, fortunately, experimental circumstances can be chosen so that the summation is non-negligible only over a finite and experimentally tractable range. We assume that we want to estimate a RIXS spectrum from $\hbar\omega_{min}$ to $\hbar\omega_{max}$ and that we want to use photoemission features extending from at least $BE_{min}$ to $BE_{max}$ in the measurement. These ranges give a PAX spectrum extending from $KE_{min} = \hbar\omega_{min}-BE_{max}$ through $KE_{max} = \hbar\omega_{max}-BE_{min}$. In order to accurately model this PAX spectrum we must keep all X-ray photon energies in Eq. \ref{eq:full_discrete_pax} that contribute non-negligibly to the PAX spectrum over this range. X-rays with energies higher than some cutoff $\hbar\omega_{+}$ give negligible contributions (a typical cutoff may be a few hundred meV above the incident X-ray energy). The lower limit of photon energies that contribute to the PAX spectrum is set through the lowest binding energy that contributes significantly to the PAX spectrum, $BE_{-}$. For example, if one is using valence photoemission features, this limit on binding energies is set by the typical restriction of significant photoemission intensity to positive binding energies.  This sets the lower limit $\hbar\omega_{-} = kE_{min}+BE_{-}$ on the photon energies of X-rays that will contribute to the PAX spectrum.

With these range limitations, Eq. \ref{eq:full_discrete_pax} is simplified to
\begin{equation}
    E\{m[kE]\} = \sum_{\hbar\omega = \hbar\omega_{-}}^{\hbar\omega_{+}} s[\hbar\omega]h[kE-\hbar\omega],
\end{equation}
which is practical to analyze for PAX. It is convenient to write this as
\begin{equation}\label{eq:discrete_pax}
    E\{\mathbf{m}\} = H\mathbf{s},
\end{equation}
where $\mathbf{m}$ is a column vector whose entries are the measured PAX spectrum, $\mathbf{s}$ is another column vector whose entries are the desired X-ray spectrum, and $H$ is a matrix such that the entries of $H\mathbf{s}$ are the same as the discrete convolution $h\ast s$.

We assume we are in a regime where shot noise is the dominant noise. In this case, the measured PAX spectrum can be approximated by a Poisson process. The probability of measuring $b$ counts for a kinetic energy bin with expected value $a$ is 
\begin{equation}
    p(b) = \frac{\exp(-a)a^b}{b!}.
\end{equation}
The probability of measuring a PAX spectrum $\mathbf{m}$ is then
\begin{equation}\label{eq:pax_probability}
    p(\mathbf{m}) = \prod_i \frac{\exp(-H\mathbf{s})_i\left(H\mathbf{s}\right)_i^{\mathbf{m}_i}}{\mathbf{m}_i!},
\end{equation}
where $\mathbf{x}_i$ denotes the ith element of $\mathbf{x}$.

 We note that a matrix equation like Eq. \ref{eq:discrete_pax} holds as a description of the expected value of a PAX spectrum even when the photoemission spectrum of the converter material is dependent on the incident photon energy. Thus, the methods we describe below can still be used to estimate a RIXS spectrum in such cases, albeit with likely less computational efficiency.

\subsection{Regularized Maximum Likelihood Estimation for Estimating RIXS with PAX}

We now describe how we estimate a ground truth RIXS spectrum given a PAX data set. We assume that we have a set of PAX spectra recorded under statistically identical conditions as well as a high accuracy measurement of the photoemission spectrum of the converter material recorded with monochromatic incident X-ray radiation. Neglecting noise of the photoemission spectrum is an acceptable approximation because the photoemission spectrum can be measured with direct photoemission which has orders of magnitude higher count rate than a PAX measurement. We estimate the ground truth RIXS spectrum from this data using regularized maximum likelihood estimation.  The maximum likelihood estimate of the ground truth RIXS spectrum is the spectrum which maximizes the probability of measuring the actually measured PAX spectrum. Regularization prevents the estimate from having finer structure than is warranted for the quality of the data.

For the probability given in Eq. \ref{eq:pax_probability}, the negative log-likelihood of $\mathbf{s}$ is
\begin{equation}
    L(\mathbf{s}) = \sum_i (H\mathbf{s})_i- \mathbf{m}_i\log((H\mathbf{s})_i)+\log(\mathbf{m}_i!).
\end{equation}
The gradient of this with respect to $\mathbf{s}$ is
\begin{equation}
    \nabla L(\mathbf{s}) = H^T\mathbf{1}-H^T\frac{\mathbf{m}}{H\mathbf{s}},
\end{equation}
where $\mathbf{1}$ is a vector where all the entries are one and with dimension such that the equation it appears in is valid, and $x^T$ denotes the transpose of $x$. Having this gradient, we can iteratively minimize the negative log likelihood (maximizing the likelihood) with the scaled gradient iteration \cite{bertero2009image}
\begin{equation}
    \hat{\mathbf{s}}^{(n+1)} = \hat{\mathbf{s}}^{(n)}-\hat{\mathbf{s}}^{(n)}\nabla L(\hat{\mathbf{s}}^{(n)}),
\end{equation}
where $\hat{\mathbf{s}}^{(n)}$ is the estimate of $\mathbf{s}$ after $n$ iterations. This gives the iteration
\begin{equation}
    \hat{\mathbf{s}}^{(n+1)} = \hat{\mathbf{s}}^{(n)}\left[\mathbf{1}-H^T\mathbf{1}+H^T\frac{\mathbf{m}}{H\mathbf{s}}\right].
\end{equation}
This is equivalent to
\begin{equation}\label{eq:deconvolve}
    \hat{s}^{(n+1)} = \hat{s}^{(n)}\left[1[\hbar\omega]-h^*\ast 1[\hbar\omega]+h^*\ast\frac{m}{h\ast \hat{s}^{(n)}}\right],
\end{equation}
where $h^*$ is the photoemission impulse response function with the order of the entries reversed and all the entries of $1[\hbar\omega]$ are one. This iteration requires a starting point, $\hat{s}^{(0)}$. For this, we used a smoothed version of the measured PAX spectrum.

If $h^*\ast 1[\hbar\omega] = 1[\hbar\omega]$, then Eq. \ref{eq:deconvolve} simplifies to the well-known Lucy-Richardson algorithm \cite{richardson1972bayesian, lucy1974iterative, shepp1982maximum}. The Lucy-Richardson algorithm and its variants have been widely used in imaging \cite{bertero2009image, dey2006richardson, starck2002deconvolution} and spectroscopy \cite{fister2007deconvolving}. For PAX, however, we often will not be able to reduce Eq. \ref{eq:deconvolve} to the Lucy-Richardson algorithm as photoemission spectra used for PAX can be non-negligible over a wide range.

It is a well-known problem that algorithms like Eq. \ref{eq:deconvolve} amplify high frequency noise when they are used without regularization \cite{white1994image, bertero2009image}. This is essentially a result of the reduction in strength of high frequency components relative to low frequency components after convolution with an extended function. Accurately inferring the pre-convolution strength of these high frequency components requires a more accurate measure of their strength in the post-convolution data then their lower frequency counterparts. 

Various regularization schemes have been proposed to avoid the amplification of high frequency noise encountered in such algorithms. This is typically achieved by enforcing some degree of smoothness of the deconvolved result. Regularization by stopping iterations after certain criteria have been met \cite{reeves1995generalized}, damping the effect of iterations that do not improve the reconstruction \cite{white1994image}, and total variation regularization \cite{dey2006richardson} have been proposed. A method of regularization well suited to our case was proposed by Fister \textit{et al.} \cite{fister2007deconvolving} for application to Lucy-Richardson deconvolution of X-ray absorption and inelastic X-ray scattering spectra. In this algorithm, the iterative deconvolution is stabilized against high frequency noise amplification by convolution with a Gaussian function after each iteration. Applying this to our algorithm gives a regularized version of Eq. \ref{eq:deconvolve},
\begin{equation}
    \hat{s}^{(n+1)} = f\ast\hat{s}^{(n)} \left[1[\hbar\omega]-h^*\ast 1[\hbar\omega]+h^*\ast\frac{m}{h\ast \hat{s}^{(n)}}\right],
\end{equation}
where $f(x)$ is a Gaussian function with unit integrated amplitude,
\begin{equation}
    f(x) = \frac{1}{\sigma\sqrt{2\pi}}\exp\left(-\frac{1}{2}(x/\sigma)^2\right).
\end{equation}
The width of the Gaussian, $\sigma$, constrains the maximum roughness of deconvolved spectra and thus sets the regularization strength (it acts as a hyperparameter in the deconvolution algorithm). Smaller regularization strengths allow for more roughness in the deconvolved spectra than larger regularization strengths.

\begin{figure}
    \centering
    \includegraphics{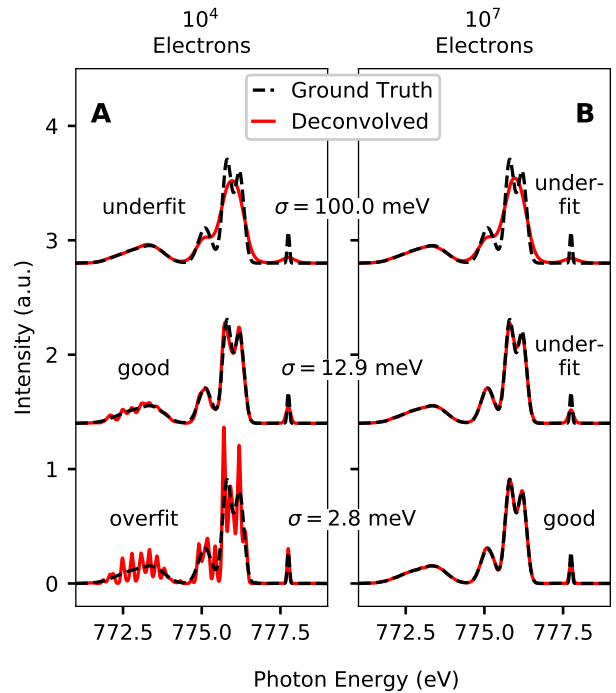}
    \caption{Effect of regularization strength on deconvolved PAX spectra. (Spectra with different regularization strengths are vertically offset for clarity.) (A) Ground truth and deconvolved spectra for different regularization strengths and $10^4$ detected electrons. (B) Same as part A, except for $10^7$ detected electrons.}
    \label{fig:effect_of_regularizer_spectra}
\end{figure}

Fig. \ref{fig:effect_of_regularizer_spectra} shows the effect of regularization strength on the deconvolved spectra for simulations using the model ground truth spectrum shown in Fig. \ref{fig:overview}A, and model Ag 3d photoemission spectrum shown in Fig. \ref{fig:overview}B. We used an energy separation of 10 meV between points for all simulations using the Ag 3d levels as a photoemission converter. Part A of Fig. \ref{fig:effect_of_regularizer_spectra} shows results for $10^4$ detected electrons, while part B shows analogous results for $10^7$ detected electrons. In each case, deconvolved and ground truth spectra are shown with regularization strength decreasing from top to bottom. As the regularization strength decreases, the deconvolved spectra attain more detail and increasingly finer structure is seen. This comes at the expense, however, of more statistical variation in the deconvolved spectra. For the case with 10$^4$ detected electrons, the deconvolved spectra accurately estimate increasingly fine spectral features with smaller regularization strengths except for the smallest regularization strength of $\sigma = 2.8$ meV. For that case, the deconvolved spectrum has fine features, but they do not accurately reflect the ground truth spectrum on this scale. In contrast, with $10^7$ detected electrons, as shown in Fig. \ref{fig:effect_of_regularizer_spectra}B, the deconvolved spectra accurately reflect the ground truth spectrum smoothed to an extent given by the particular regularization strength even for the smallest regularization strength shown. Thus, the optimal regularization strength is smaller for $10^7$ detected electrons than for $10^4$ detected electrons. More generally, the best regularization strength decreases with increasing numbers of detected electrons, but also depends on the converter material photoemission spectrum and the ground truth X-ray spectrum.

\begin{figure*}
    \centering
    \includegraphics[]{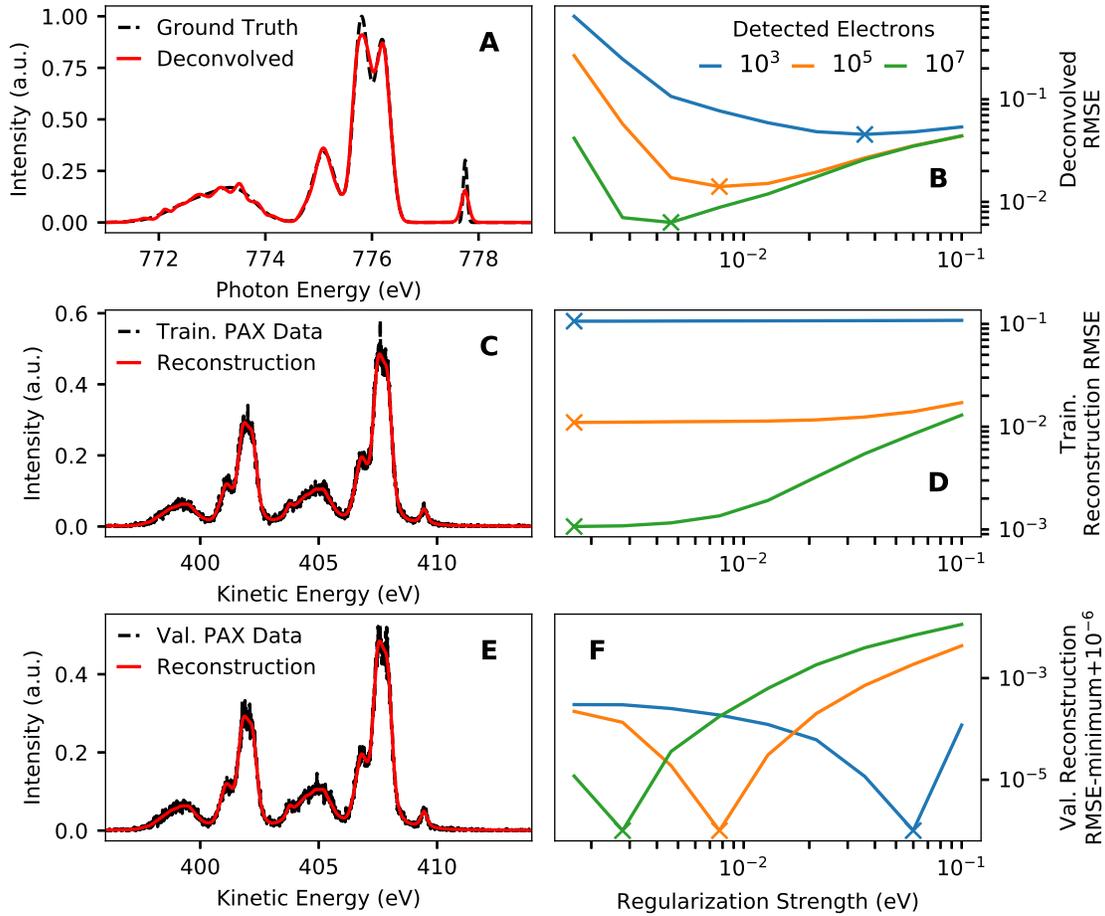}
    \caption{Illustration of how the quality of reconstruction of PAX spectra from deconvolved spectra can be used to estimate the optimal regularization strength. The panels on the left show example simulated data with $10^5$ detected electrons and a regularization strength of 7.7 meV. (A) example ground truth spectrum and an estimate of it obtained by deconvolving simulated data. (B) corresponding Root Mean Squared Error (RMSE) of the deconvolved spectrum as a function of the regularization strength and the number of detected electrons. (C) PAX spectrum obtained by averaging a training set of data (data that was used in deconvolution) and its reconstruction from the deconvolved result. (D) corresponding RMSE of the reconstruction of the training data for the same parameters as A. (E) PAX spectrum obtained by averaging a validation set of data (data that was not used in deconvolution) and its reconstruction from the deconvolved result. (F) corresponding RMSE of the validation data for the same parameters as A. The data of panel (F) is shown with the minimum of each curve subtracted and a small offset added in order to highlight the locations of minima. Since the validation reconstruction error is minimized at similar regularization strengths as the deconvolved error, we can estimate the optimal regularization strength from the validation reconstruction error.}
    \label{fig:effect_of_regularizer_quant}
\end{figure*}

\subsection{Estimating the Optimal Regularization Strength}

We now show how we can estimate the optimal regularization strength. Fister \emph{et al.} \cite{fister2007deconvolving} proposed one method of choosing the regularization based on the smallest expected feature in the ground truth spectrum. Since we do not always know this before measurements, we use a more general procedure here. We define the optimal regularization strength as that which minimizes the Root Mean Squared Error (RMSE) of a deconvolved spectrum with respect to the ground truth it estimates,
\begin{equation}
    \hat{\sigma} = \argmin\{\sqrt{||\hat{\mathbf{s}}-\mathbf{s}||^2}\}.
\end{equation}
Fig. \ref{fig:effect_of_regularizer_quant} shows the dependence of different statistics on the regularization strength and number of detected electrons. These simulations were carried out using the Ag 3d photoemission spectrum shown in Fig. \ref{fig:overview}C as a model photoemission converter and the model RIXS spectrum of Fig. \ref{fig:overview}A as a ground truth spectrum to estimate. The results shown in Fig. \ref{fig:effect_of_regularizer_quant} can be used to assess the potential of the different statistics in estimating the optimal regularization strength. 

Fig. \ref{fig:effect_of_regularizer_quant}A shows an example ground truth spectrum and a deconvolved estimate of it from simulated PAX data. Calculating the RMSE of such deconvolved spectra as a function of the regularization strength and number of detected electrons results in the curves shown in Fig. \ref{fig:effect_of_regularizer_quant}B. This deconvolved RMSE decreases with increasing regularization strength down to the minimum at the optimal regularization strength, where it then gradually increases with increasing regularization strength. The optimal regularization strength is smaller for higher numbers of detected electrons. Unfortunately, this deconvolved RMSE is not experimentally accessible as the ground truth spectrum is generally unknown.

Instead of trying to determine the optimal regularization strength through direct assessment of deconvolved error, we can assess how well our model reconstructs PAX spectra from deconvolved spectra as a proxy for the accuracy of deconvolved spectra. In other words, we can compare the convolution of a deconvolved spectrum with the corresponding photoemission impulse response function to recorded PAX data. If the deconvolution is perfect and there is no noise, these spectra should be the same. It has been shown in previous deconvolution studies that such a comparison can allow one to estimate the optimal regularization strength \cite{reeves1992cross, wahba1990optimal}. In making these assessments, we distinguish between two cases. In the first case, we compare a reconstruction of PAX data to a training PAX spectrum (the same spectrum that was used as input for deconvolution). In the second case, we compare a reconstruction of PAX data to a validation PAX spectrum (a spectrum recorded in statistically identical conditions as the training spectrum, but that was not used as part of that deconvolution).

Fig. \ref{fig:effect_of_regularizer_quant}C shows a comparison of a reconstruction of a PAX spectrum to a training PAX spectrum. Fig. \ref{fig:effect_of_regularizer_quant}D shows the RMSE of such reconstructions as a function of regularization strength and number of detected electrons. Unfortunately, as seen there, this statistic only decreases with decreasing regularization strength and is not minimized at the same locations as the deconvolved RMSE. Thus we cannot use the minimum of this statistic to estimate the optimal regularization parameter. Deconvolutions with too small regularization strengths closely fit fine features of the recorded PAX spectrum even though fitting so closely is not warranted given the noise in the data. This is an example of overfitting. To avoid this problem, we can assess the performance of the deconvolution on data that was not used as input to the deconvolution \cite{james2013introduction}.

Fig. \ref{fig:effect_of_regularizer_quant}E compares the PAX reconstruction to a validation PAX spectrum. Fig. \ref{fig:effect_of_regularizer_quant}F shows the dependence of the RMSE of the reconstruction with respect to the validation PAX spectrum on the regularization strength and the number of detected electrons. In each case, this statistic is minimized near the optimal regularization strengths. Unlike the RMSE of deconvolved data shown in Fig. \ref{fig:effect_of_regularizer_quant}B, this statistic is experimentally accessible, and thus we use it to estimate the optimal regularization strength.

\begin{figure}
    \centering
    \includegraphics[]{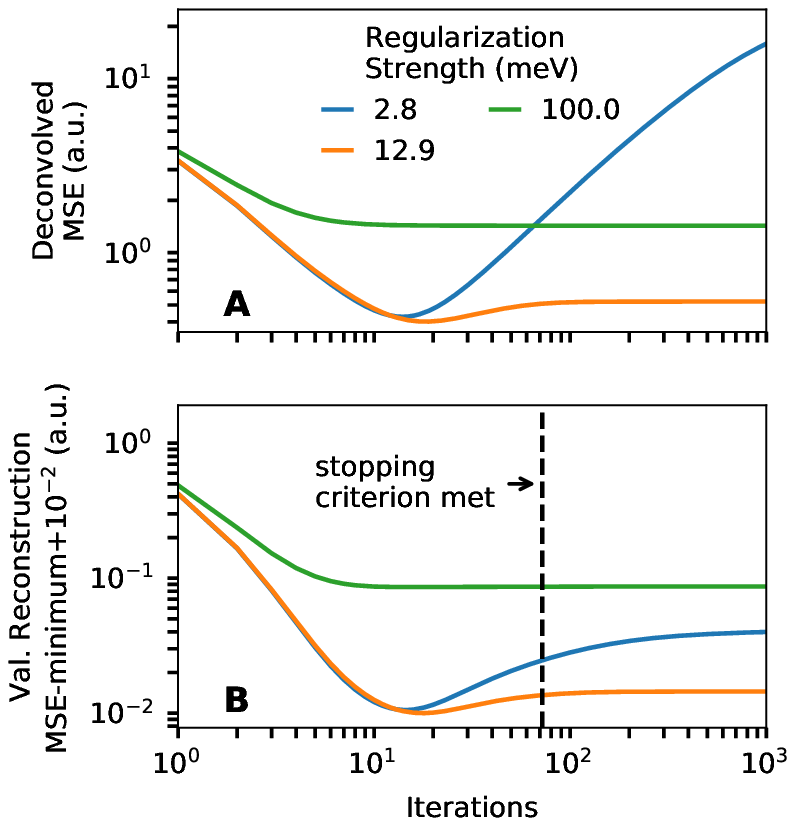}
    \caption{Illustration of stopping criterion for the same data set as Fig. \ref{fig:effect_of_regularizer_spectra}A. (A) shows the dependence of the deconvolved mean squared error (MSE) on iteration number and regularization strength, while (B) shows that for the mean squared error of the reconstruction of validation data. The stopping criterion is four times the number of iterations required for the validation error to reach a minimum with the smallest tested regularization strength.}
    \label{fig:stopping}
\end{figure}

\subsection{Stopping Criterion}

\begin{figure}[htb]
    \centering
    \includegraphics{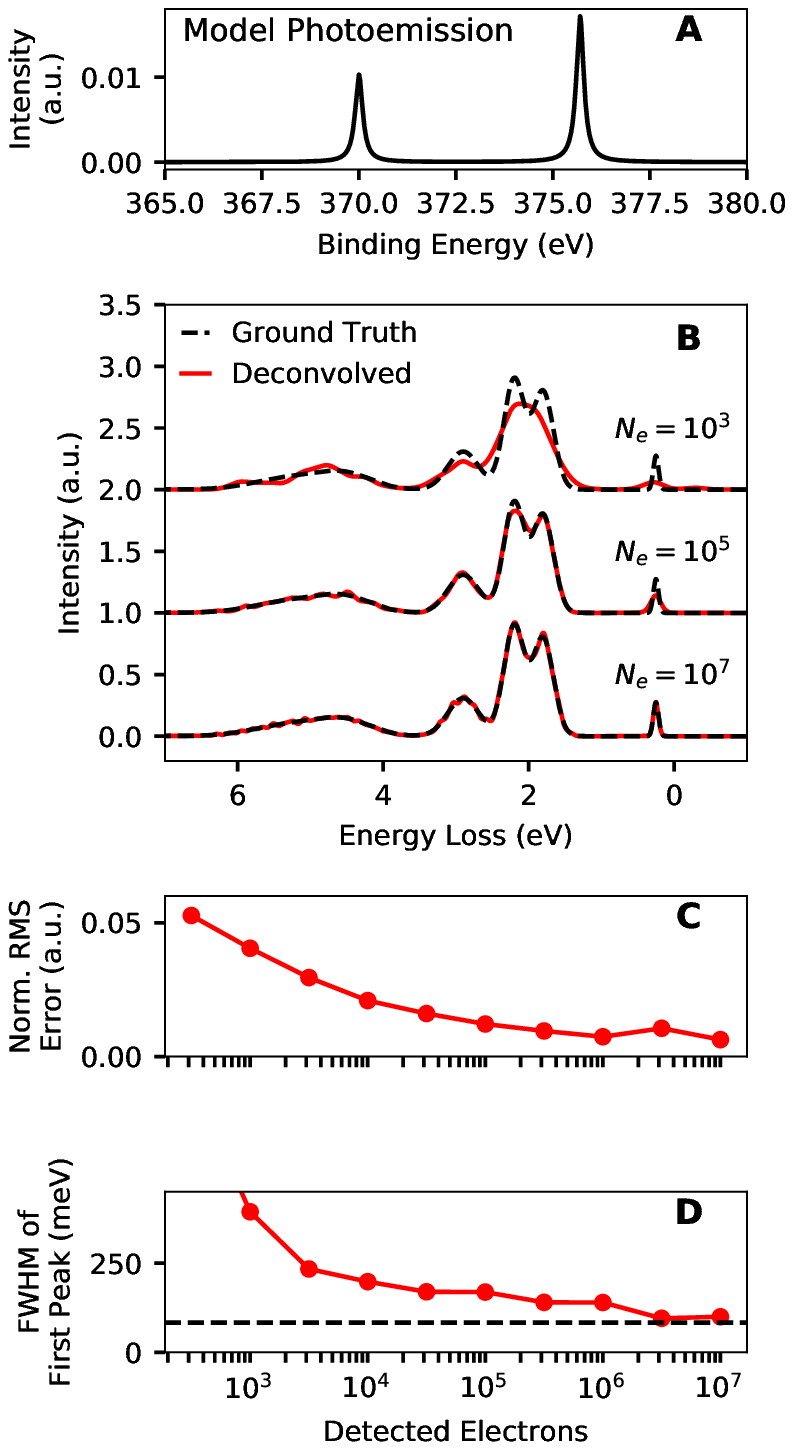}
    \caption{Performance of PAX in estimating a model RIXS spectrum using a model Ag 3d photoemission converter. (A) Model Ag 3d photoemission spectrum with 233 meV FWHM Lorentzian peaks\cite{panaccione2005high}. (B) RIXS spectra estimated through deconvolution of simulated PAX data for $10^3$, $10^5$ and $10^7$ detected electrons. For each case, the ground truth spectrum is also shown. (B) Normalized RMS error of deconvolved spectra versus the number of detected electrons. (C) FWHM of the lowest energy loss peak of the deconvolved RIXS spectrum versus the number of detected electrons. The dashed line shows the ground truth value.}
    \label{fig:performance1}
\end{figure}

To perform iterative deconvolution, it is necessary to decide when a sufficient number of iterations have been completed. To choose such a stopping criterion, we recognized that (1) we wanted to perform sufficient iterations for the deconvolution with the optimal regularization strength to converge to near its asymptotic value and (2) we wanted the number of iterations to be high enough for the optimal regularization strength to be closely approximated by that which minimizes the validation reconstruction RMSE.

 Fig. \ref{fig:stopping} illustrates the stopping criterion we used for this study. This uses data simulated with the same conditions as Fig. \ref{fig:effect_of_regularizer_spectra}A. For smaller regularization strengths, the deconvolved and validation reconstruction mean squared errors reach minima before increasing again. For larger regularization strengths, the deconvolved and validation reconstruction mean squared error decrease to near their asymptotic value quickly without overshooting. We see that, for this data, to fulfill the above conditions, it is necessary to complete iterations at least a few times more than that where the errors for the smallest regularization strength are minimized. In our case we chose to conduct iterations at least equal to four times that where the validation RMSE for the smallest regularization strength reaches a minimum. We note that it can still be beneficial to do more iterations where possible to better fulfill the above priorities. 
 
 We have now finished describing an algorithm for estimating a ground truth X-ray spectrum from PAX data. We summarize this procedure in Algorithms \ref{alg:rLR} and \ref{alg:estimateSigma}. The code used to perform the deconvolution analysis described in this report can be found at
github.com/dhigley6/PAX2.

\begin{algorithm}[htb]
\SetAlgoNoLine
\KwIn{m: measured PAX spectrum, h: PAX impulse response function (measured converter photoemission spectrum as a function of negative binding energy), $\sigma$: regularization strength}
\KwOut{$\hat{s}$: regularized maximum likelihood estimate of target X-ray spectrum}
$f(x) \gets \frac{1}{\sigma\sqrt{2\pi}}\exp\left(-\frac{1}{2}(x/\sigma)^2\right)$ \;
$n \gets 0$ \;
$\hat{s}^{(0)} \gets f\ast m$ \;
\While{stopping criterion not met}{
  \hspace{10pt}$\hat{s}^{(n+1)}\gets f \ast \hat{s}^{(n)} \left[1[\hbar\omega]-h^*\ast 1[\hbar\omega]+h^*\ast\frac{m}{h\ast \hat{s}^{(n)}}\right]$\;
  \hspace{10pt}$n \gets n+1$ \;
}
$\hat{s} \gets \hat{s}^{(n)}$ \;
\caption{Regularized Deconvolution of PAX Data}
\label{alg:rLR}
\end{algorithm}

\begin{algorithm}[htb]
\SetAlgoNoLine
\KwIn{M: set of $k$ measured PAX spectra with the nth spectrum labeled $m_n$, h: PAX impulse response function (measured converter photoemission spectrum as a function of negative binding energy), $\mathcal{S}$: set of regularization strengths to choose among, $D(m, h, \sigma)$: regularized deconvolution of m with impulse response function $h$ and regularization strength $\sigma$, as described in Algorithm \ref{alg:rLR}}
\KwOut{$\hat{\sigma}$: estimate of regularization strength which minimizes the squared difference between the deconvolved and ground truth X-ray spectrum}
\For{$\sigma \in \mathcal{S}$}{
  \hspace{10pt}\For{$m_n \in M$}{
    \hspace{20pt}$m_{-n} = \frac{1}{k-1}\sum_{m_i, i \neq n}m_i$ \;
    \hspace{20pt}$\hat{s}_{-1} \gets D(m_{-1}, h, \sigma)$ \;
    \hspace{10pt}}
  \hspace{10pt}$CV(\sigma) \gets \frac{1}{k}\sum_{i=1}^k ||\hat{s}_{-i}\ast h-m_i||^2$ \;
}
$\hat{\sigma} \gets argmin(CV)$
\caption{Estimate Optimal Regularization Strength for Regularized Deconvolution}
\tcc{After estimating an optimal regularization parameter, run Algorithm 1 with this regularization parameter to get the deconvolved spectrum}
\label{alg:estimateSigma}
\end{algorithm}

\section{Performance of PAX}\label{sec:results}

Now that we have described an algorithm for analyzing PAX spectra, in this section we evaluate the performance of this algorithm on simulated data. We simulated PAX measurements using the Ag 3d photoemission lines to estimate features with structure on a 100 meV scale as well as a sharp Fermi edge such as seen in Au to estimate finer features.

Fig. \ref{fig:performance1} shows the performance of PAX with the Ag 3d lines as a photoemission converter (A) in estimating the model RIXS spectrum shown in Fig. \ref{fig:overview}A. Part B shows ground truth and deconvolved X-ray spectra for the number of detected electrons increasing from top to bottom. As the number of detected electrons increases finer details of the X-ray spectrum are accurately estimated. Features with a few hundred meV width are already estimated well with $\approx 10^5$ detected electrons, and the width of well estimated features decreases further with increasing number of detected electrons. The RMSE of the deconvolved spectrum decreases as the number of detected electrons increases, as shown in Fig. \ref{fig:performance1}C. Fig. \ref{fig:performance1}D quantifies the ability of the method to accurately estimate fine features through the FWHM of the lowest energy loss peak of the deconvolved spectrum as a function of the number of detected electrons. This FWHM decreases as the number of detected electrons increases, and, after $10^{6.5}$ electrons have been detected, the width of the feature in the deconvolved spectrum is within 10 percent of its true width. We note that the integrated intensity of this first loss feature is only 2.4 percent of the total integrated intensity of the entire model RIXS spectrum. Thus one only needs to detect less than $10^4$ photoelectrons that were emitted from RIXS photons originating from this feature in order to accurately estimate the shape of this feature. We accurately estimated this feature despite it being much sharper than the the 233 meV FWHM widths of the Ag 3d photoemission peaks of the model impulse response function.

\begin{figure}
    \centering
    \includegraphics{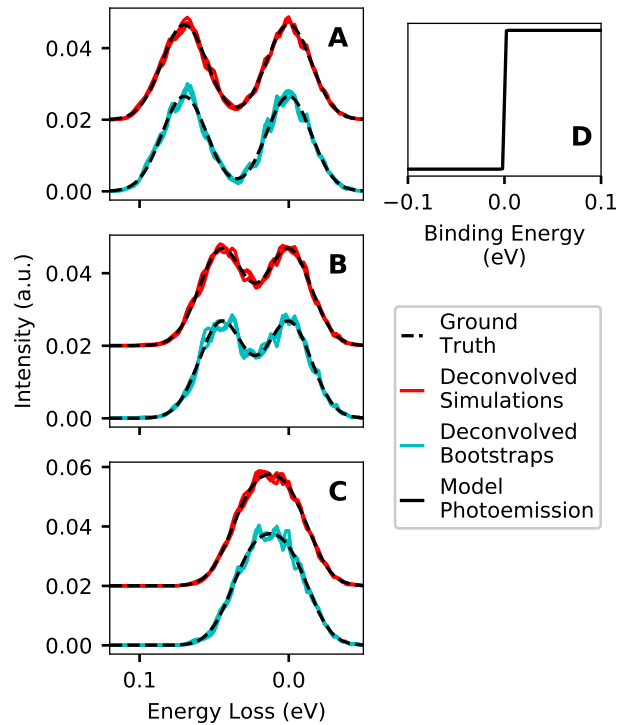}
    \caption{Performance of PAX in estimating model doublet RIXS spectra using a sharp Fermi edge photoemission converter. For each case, $\approx 10^7$ electrons were detected from photoemission within 0.4 eV of the Fermi level. The doublets consisted of two Gaussian peaks with 100 meV FWHM each and variable peak separation. The shape of the Fermi edge was determined by a Fermi-Dirac distribution at 4K (boiling point of helium). For each case, we show three spectra deconvolved from independent PAX simulations and three deconvolutions of PAX spectra estimated using bootstraps of a single simulated PAX data set. (A) doublet spectra with 70 meV peak separation. (B) doublet spectra with 45 meV peak separation. (C) doublet spectra with 25 meV peak separation. (D) Model photoemission spectrum.}
    \label{fig:doublet}
\end{figure}

Fig. \ref{fig:doublet} shows an analysis of the performance of PAX in estimating spectra with structure on scales finer than 100 meV by using a sharp Fermi edge. For this analysis we used an energy separation of 2 meV between points. The photoemission spectrum was modeled with a constant density of states near the Fermi level and a temperature of 4K (boiling point of He). PAX spectra were simulated for $10^7$ electrons detected from photoemission within 0.4 eV of the Fermi level and variable separation of the X-ray doublet peaks. For each panel of Fig. \ref{fig:doublet}, we compare three deconvolved simulated PAX spectra to the ground truth spectrum. By deconvolving more than one simulated PAX spectrum we get an indication of whether features in the deconvolved spectrum are reproducible with the experimental accuracy. This check has the drawback that not all of the data is used in forming a single deconvolved spectrum. In order to get an idea of whether features are reproducible without dividing the data into subsets, we can use the bootstrap method \cite{efron1986bootstrap, james2013introduction}. In this method, we approximate the distribution of PAX spectra that we would obtain from averaging $n$ recorded PAX spectra that compose a PAX data set (in this case $n=1000$). We do this by forming bootstrapped PAX spectra from the data by sampling $n$ times from the $n$ recorded PAX spectra and averaging the results. The sampling is done with replacement so that some samples may be chosen more than once and others not at all. We can then tell if a feature is unlikely to be a random occurrence by confirming whether it consistently occurs in a large fraction of the deconvolved bootstrapped PAX spectra. For clarity, only three such spectra are shown in Fig. \ref{fig:doublet}, but to make scientific conclusions one should look at several times more spectra \cite{efron1986bootstrap}.

As the peak separation decreases from top to bottom in Fig. \ref{fig:doublet}, our ability to tell that the ground truth spectrum is two peaks rather than a single peak decreases. With the 70 meV peak separation (A) as well as the 45 meV peak separation (B), it is clear that the deconvolved spectra are not well represented by a single peak, and this is reproducibly the case between independent measurements or bootstrapped measurements. In contrast, for the 25 meV peak separation (C), it is no longer clear that the ground truth spectrum consists of more than one Gaussian peak, and a measurement with better statistics would be required to tell this.

\section{Conclusions and Outlook}\label{sec:conclusions}
We have demonstrated the potential of PAX for measurement of RIXS spectra. Using the Ag 3d levels as a photoemission converter for PAX, few hundred meV FWHM features can be accurately estimated with $10^5$ detected electrons. Even finer features can be accurately estimated with higher numbers of detected electrons. Details with aspects much smaller than 100 meV could be estimated using a sharp Fermi edge photoemssion converter, albeit at the expense of reduced conversion efficiency of RIXS photons to electrons.

In this report, we proposed and tested one algorithm which can be used to estimate a RIXS spectrum from measured PAX data. Our algorithm is simple and closely linked to the classic Lucy-Richardson algorithm. Recently, however, more sophisticated algorithms have shown much promise on achieving more accurate deconvolution results and reducing computational time \cite{ikoma2018convex, zhang2017learning}. Applying such techniques to PAX could improve on the performance described here. In addition, using uncertainty quantification methods would enable more robust interpretation of X-ray spectra estimated from PAX data. This has been done for similar problems by assessing the sensitivity of deconvolved spectra to artificially added noise \cite{fister2007deconvolving} as well as more sophisticated methods \cite{kaipio2006statistical}. 

Additional experimental development could also push the capability of PAX further. The model photoemission converters highlighted here were chosen based on a survey of literature photoemission data. A systematic investigation of other materials may provide photoemission features better suited for PAX measurements. Finally, the PAX measurement method could be applied to other situations where high signal-to-noise ratio estimates of X-ray spectra are desired without using traditional grating-based technology. An example could be transmissive soft X-ray spectrometers.

\begin{acknowledgments}
We acknowledge A. Owen for a helpful conversation. Use of the Linac Coherent Light Source (LCLS), SLAC National Accelerator Laboratory, is supported by the U.S. Department of Energy, Office of Science, Office of Basic Energy Sciences under Contract No. DE-AC02-76SF00515. Use of the Stanford Synchrotron Radiation Lightsource, SLAC National Accelerator Laboratory, is supported by the U.S. Department of Energy, Office of Science, Office of Basic Energy Sciences under Contract No. DE-AC02-76SF00515.
\end{acknowledgments}

\bibliography{pax_refs}

\end{document}